\documentclass[10pt,journal,final]{IEEEtran}
\IEEEoverridecommandlockouts
\usepackage{epsfig,latexsym,amssymb,amsmath,graphics,color}
\usepackage{graphicx,subfigure}
\newtheorem{prop}{Proposition}

\newtheorem{cor}{Corollary}

\newtheorem{lm}{Lemma}

\newtheorem{thm}{Theorem}

\newcommand{\be}{\begin{eqnarray}}
\newcommand{\ee}{\end{eqnarray}}
\newcommand{\benn}{\begin{eqnarray*}}
\newcommand{\eenn}{\end{eqnarray*}}
\def\IR{\rm I \kern-0.20em R}

\newcommand{\bthm}{\begin{thm}}
\newcommand{\ethm}{\end{thm}}

\newcommand{\bcor}{\begin{cor}}
\newcommand{\ecor}{\end{cor}}
\newcommand{\bprop}{\begin{prop}}
\newcommand{\eprop}{\end{prop}}
\newcommand{\blm}{\begin{lm}}
\newcommand{\elm}{\end{lm}}
\newcommand{\beq}{\begin{equation}}
\newcommand{\eeq}{\end{equation}}
\newcommand{\ber}{\begin{eqnarray}}
\newcommand{\eer}{\end{eqnarray}}

\newcommand{\bproof}{\begin{proof}}
\newcommand{\eproof}{\end{proof}}

\newcommand{\bit}{\begin{itemize}}
\newcommand{\eit}{\end{itemize}}
\newcommand{\ben}{\begin{enumerate}}
\newcommand{\een}{\end{enumerate}}
\newcommand{\bdesc}{\begin{description}}
\newcommand{\edesc}{\end{description}}
\newcommand{\beqarrn}{\begin{eqnarray*}}
\newcommand{\eeqarrn}{\end{eqnarray*}}
\newcommand{\bproofof}{\begin{proofof}}
\newcommand{\eproofof}{\end{proofof}}
\newenvironment{rem}{\begin{trivlist}\item[]{\bf
Remark:}\hspace{4mm}}{\end{trivlist}}
\newcommand{\brem}{\begin{rem}}
\newcommand{\erem}{\end{rem}}
\newenvironment{rems}{\begin{trivlist}\item[]{\bf
Remarks}\begin{itemize}}{\end{itemize}\end{trivlist}}
\newcommand{\brems}{\begin{rems}}
\newcommand{\erems}{\end{rems}}
\newtheorem{fact}{Fact}
\newcommand{\bfact}{\begin{fact}}
\newcommand{\efact}{\end{fact}}
\newtheorem{examp}{Example}
\newcommand{\bexamp}{\begin{examp}\rm}
\newcommand{\eexamp}{\end{examp}}
\newtheorem{defn}{Definition}
\newcommand{\bdefn}{\begin{defn}\rm}
\newcommand{\edefn}{\end{defn}}

\newtheorem{alg}{Algorithm}
\newcommand{\balg}{\begin{alg}}
\newcommand{\ealg}{\end{alg}}

\newtheorem{prob}{Problem}
\newcommand{\bprob}{\begin{prob}}
\newcommand{\eprob}{\end{prob}}

\newcommand{\bvtm}{\begin{verbatim}}
\newcommand{\bfig}{\begin{figure}}
\newcommand{\efig}{\end{figure}}
\newcommand{\bcen}{\begin{center}}
\newcommand{\ecen}{\end{center}}

\def\beqa{\begin{eqnarray}}
\def\eeqa{\end{eqnarray}}

\long\def\comment#1{}

\def \n2{{N_0 \over 2}}

\def \h5{\hspace{0.5in}}

\begin{document}
\title{Sunlight Enabled Vehicle Detection by LED Street Lights}
\author{Weicheng Xue, Shangbin Li, and Zhengyuan Xu
\thanks{This work was supported by Key Program of National Natural Science Foundation of China (Grant No. 61631018), National Natural Science Foundation of China (Grant No. 61501420), Key Research Program of Frontier Sciences of CAS (Grant No. QYZDY-SSW-JSC003).}
\thanks{W. Xue, S. Li and Z. Xu are with Key Laboratory of Wireless-Optical Communications, Chinese Academy of Sciences, University of Science and Technology of China, Hefei, Anhui 230027, China. Email: \{shbli, xuzy\}@ustc.edu.cn.}}

\maketitle
\thispagestyle{empty}
\pagestyle{empty}
\begin{abstract}
We propose and demonstrate a preliminary traffic sensing system based on the widely distributed LED street lights. The system utilizes and discriminates the photoelectric responses of the LEDs to sunlight when a vehicle moves through the LEDs¡¯ field of view (FOV) aiming at the road. A data vector is constructed from the consecutively collected time samples of a moving observation window, and a support vector machine (SVM) based learning algorithm is subsequently developed to classify the presence of a vehicle. Finally, we build a simulated platform to experimentally evaluate the performance of the vehicle detection algorithm.
\end{abstract}

\begin{IEEEkeywords}
Visible light sensing; Passive sensing; Internet of Things; Vehicle detection
\end{IEEEkeywords}
\noindent
\section{Introduction}\label{sec1}

Vehicle detection has become an indispensable part of the intelligent transportation system. Through monitoring and collection of traffic flow data, it is possible to analyze the distributions of the traffic flow density and speed. With the development of the visible light communication (VLC) technology \cite{Mao}, much attention has been devoted to constructing various application scenarios of joint lighting and VLC. 
Visible light sensing (VLS) has also intrigued increasing interests. The VLS system can directly utilize the illumination LEDs as the sensors \cite{Li, Licc}, and analyzes the response of the LED receiver to the ambient light, thereby inferring the relevant information of the environment.

Here, we propose a novel approach of using LED street lights for vehicle detection. The road is illuminated by sunlight, and LEDs respond to the reflected light. With a simple amplification circuit designed to connect to the power supply of the street lights, the LED array inside the street lamps can be used as a light detector to sense the variation of the ambient light intensity caused by the passing vehicles.
Furthermore, it is demonstrated the support vector machine (SVM) algorithms could help to decrease probabilities of the miss detection and false alarm.

\section{System Model}\label{sec2}

Most of the VLS systems contain three elements: light sources, objects, and receivers \cite {Yang}. Light sources are divided into two categories according to whether or not it can modulate information: the active source or passive source. Objects are also divided into two categories according to whether it contain receivers \cite {Zuniga}.

Here the vehicle flow detection system we proposed is a full-passive (\textit{passive source, passive object}) system. It uses sunlight as the light source, and only adopts LED street lights as detectors. Part of the light is reflected into the LED street lamps. The LED array inside the street lamps receives the reflected light and generates electrical response signal. Through the analysis of the response signals, the system can output traffic flow in the area covered by the street lights.

As shown in Fig. 1, we establish a schematic model of LED street light based vehicle detection and build a scaled-down experimental platform in the laboratory.

\begin{figure}[!htbp]
	\centering
	\centerline{\includegraphics[width=0.35\columnwidth,angle=-90]{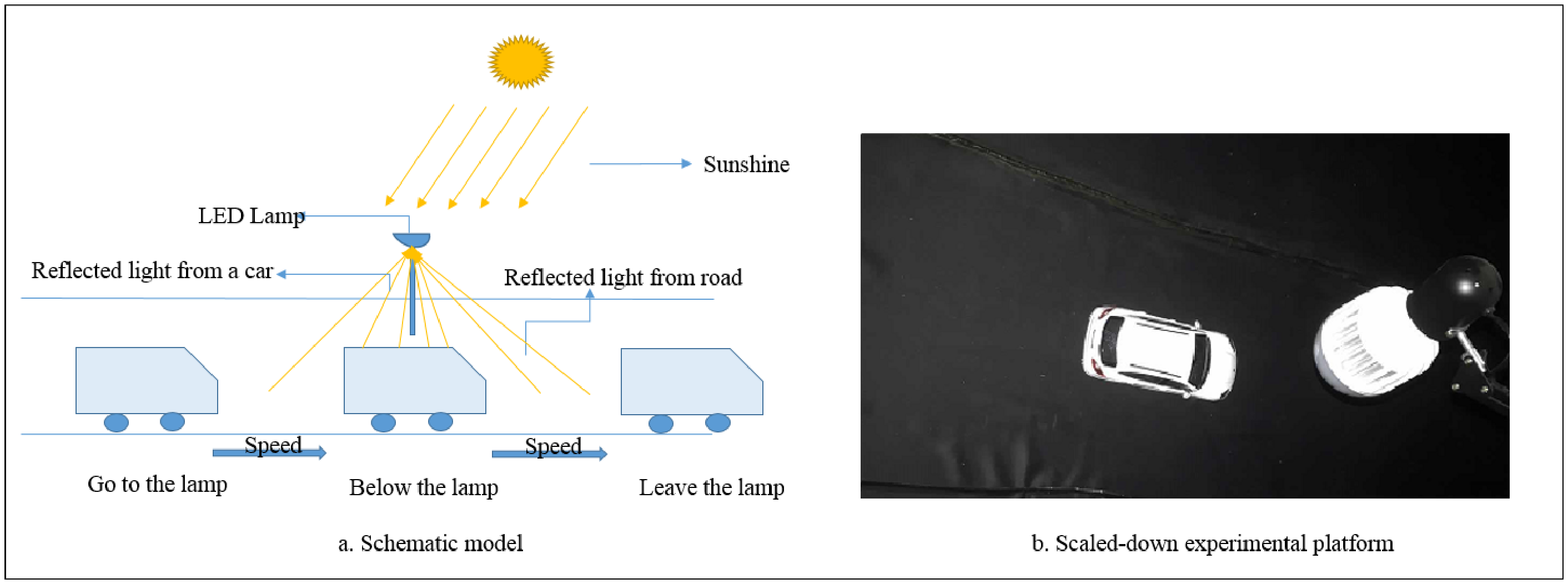}}
	\caption{The schematic model and the scaled-down experimental LED street lighting based vehicle detection.}\label{J3_Po_2}
\end{figure}

In the sunlight conditions, the ambient light enters the LED street lamps through the reflection of the road surface, so that the LED street lights can generate a response current during the day even though they do not emit light. When a car enters the sensing region of individual street lights, it interferes with the light environment, and the interference signal can be captured by the LED. Utilizing existing street lighting systems to achieve visible light sensing can help to realize smart traffic.

\section{LED-based VLS}\label{sec3}

Recently, with the development of visible light communications, much attention has been paid to LEDs as the VLC receivers and VLS sensors \cite{Dietz, Li, Licc}.
The outdoor lighting system extensively uses LED street lights as light sources, which provides the natural hardware basis of sensors, thereby reducing the cost of the sensing system.

\subsection{LED Street Light Sensing}\label{sec3.1}

The photodetector is an important part of visible light sensing. It can convert optical signals into electrical signals.
We have tested the relationship between the LED response current and the illuminance over the surface of the LED array.
As shown in Fig. 2, when the reverse current of the LED is not saturated, the LED response current is approximately proportional to the incident light intensity.

\begin{figure}[!htbp]
	\setlength{\abovecaptionskip}{-0.0cm}
	\setlength{\belowcaptionskip}{-0.0cm}
	\centering
	\centerline{\includegraphics[width=0.8\columnwidth]{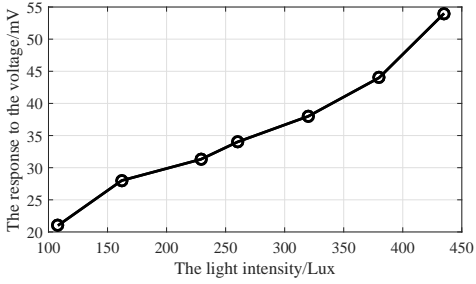}}
	\caption{Measured response voltage versus injected light intensity.}\label{J3_Po_2}
\end{figure}

Therefore, we can obtain the information of the ambient light intensity through the LED's response current, and then use the change of the ambient light intensity information to infer the state of the traffic flow on the road.

\subsection{Single Sensing Unit}\label{sec3.2}
We have developed a sensing unit that consists of a $3\times9$ LED chip array from a street light, a fish-eye lens and a sampling circuit.

\subsubsection{Signal Smoothing}\label{sec3.2.1}
The raw readout of the response current signal is not smooth. The amplitude of the noise is very small relative to the outdoor light intensity. So we use the averaged data to analyze the signal state of traffic flow.

\subsubsection{Influence of Ambient Light Intensity and Cars' Speed}\label{sec3.2.2}
The LED array can be used as a light sensor because it produces different electrical responses when the input light intensity is different.
 This means that LED lights have different perceptions of traffic signals during different time periods of a day. Also, when a car passes the covered road area of a street light, the waveform of the response current signal generated is related to the speed of the car. The faster the car speed, the smaller the pulse width of the signal.
 Figure 3(a) and Fig. 3(b) respectively show the response signals due to the car movement through street lights under different ambient light intensities and at different vehicle speeds.

\begin{figure}[!ht]
\centering
\subfigure[]{\includegraphics[width=0.45\columnwidth,angle=-0]{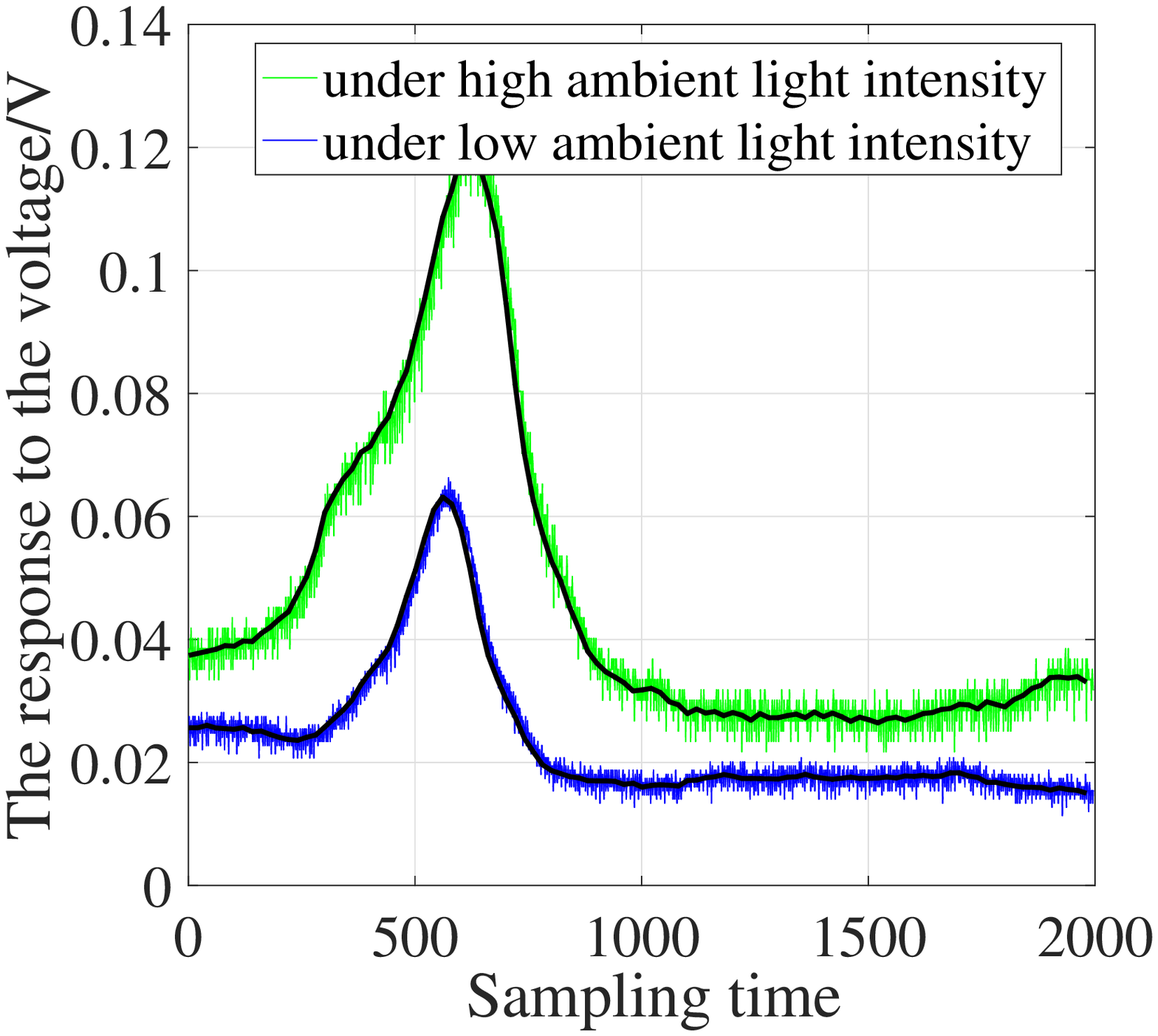}}
\subfigure[]{\includegraphics[width=0.45\columnwidth,angle=-0]{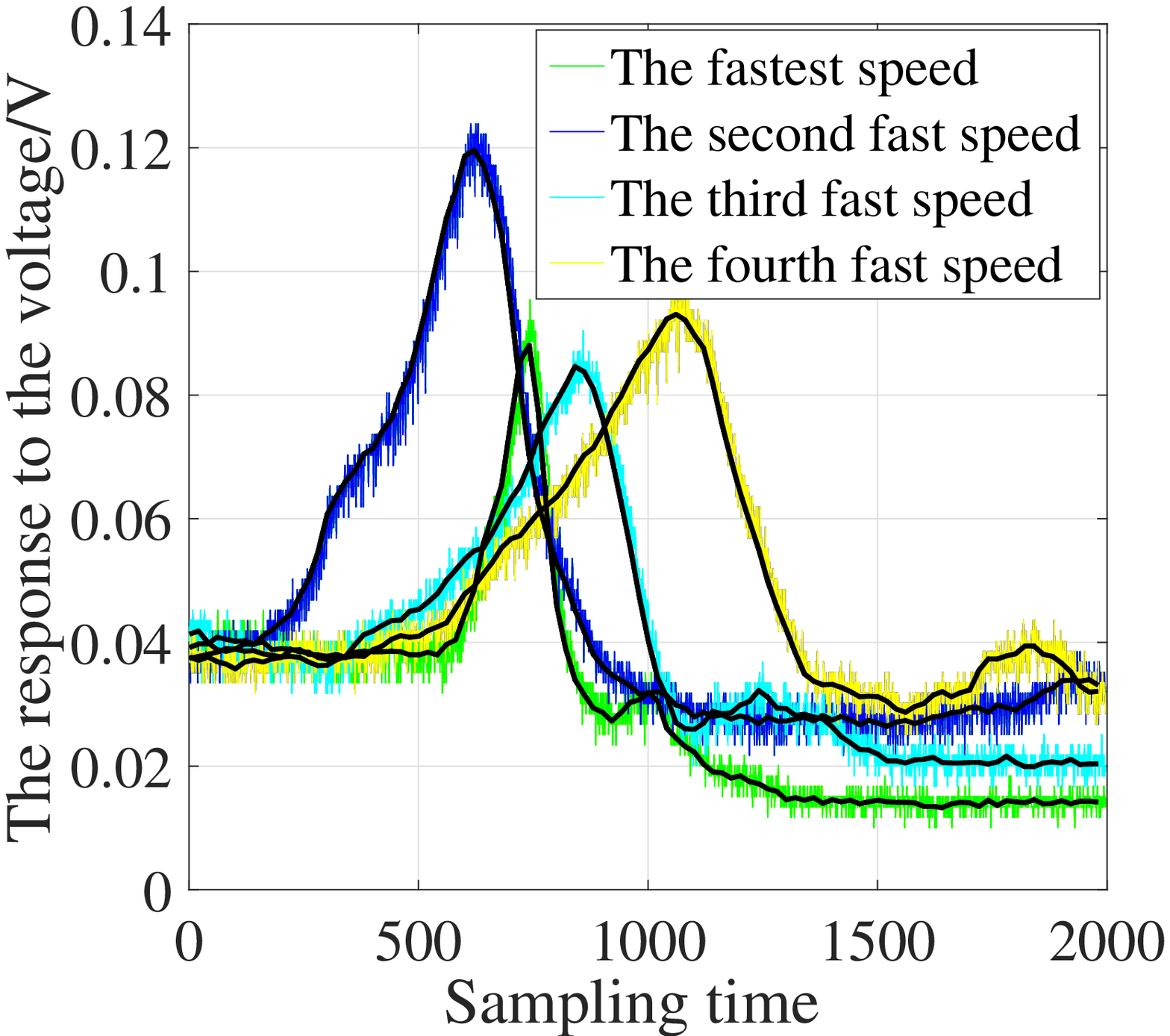}}
\caption{(a) Signal waveform under different illuminance; (b) Signal waveform at different speeds.}
\end{figure}

\section{Vehicle Detection}\label{sec4}
Because the illuminance caused by the sunlight and the speed of cars might change, the threshold based detection is not reliable.
We implement data processing and develop a support vector machine (SVM) algorithm \cite{Xia} to detect the vehicle moving under the LED street lights.

\subsection{Sensing Value Vectors}\label{sec4.1}
We denote the sensing value at sampling time $t$ as $x(t)$ and we use a moving window to obtain a time-varying vector $\mathbf{x}_t$. The real-time window truncates measured data at continuous sampling times, and the last sampling time is the current time. In other words, at sampling time $t$, the input vector $\mathbf{x}_t$ is
\begin{equation}\label{split}
\mathbf{x}_t=\Big[x\left(t-T+1\right),x\left(t-T+2\right),\cdots,x\left(t\right)\Big]^{T},
\end{equation}
where $T$ is the number of sampling times. Figure 4 illustrates the process of moving window to construct the data vector.

\begin{figure}[!htbp]
\setlength{\abovecaptionskip}{-0.0cm}
\setlength{\belowcaptionskip}{-0.0cm}
\centering
\includegraphics[width=1.0\columnwidth,angle=0]{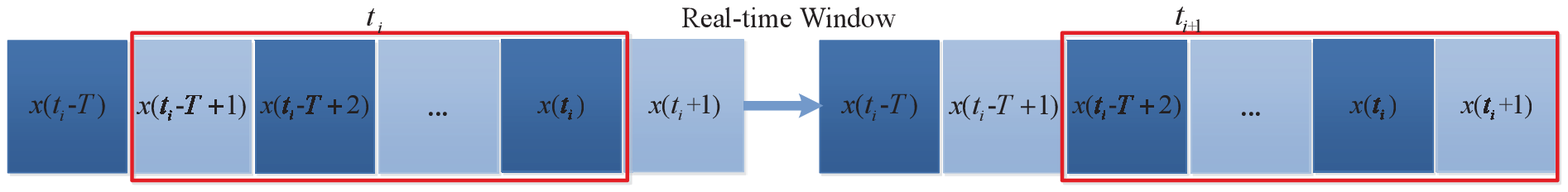}
\caption{Data vector generation by moving window.}
\end{figure}

\begin{figure*}[!htbp]
	\setlength{\abovecaptionskip}{-0.0cm}
	\setlength{\belowcaptionskip}{-0.0cm}
	\centering
	\centerline{\includegraphics[width=1.8\columnwidth]{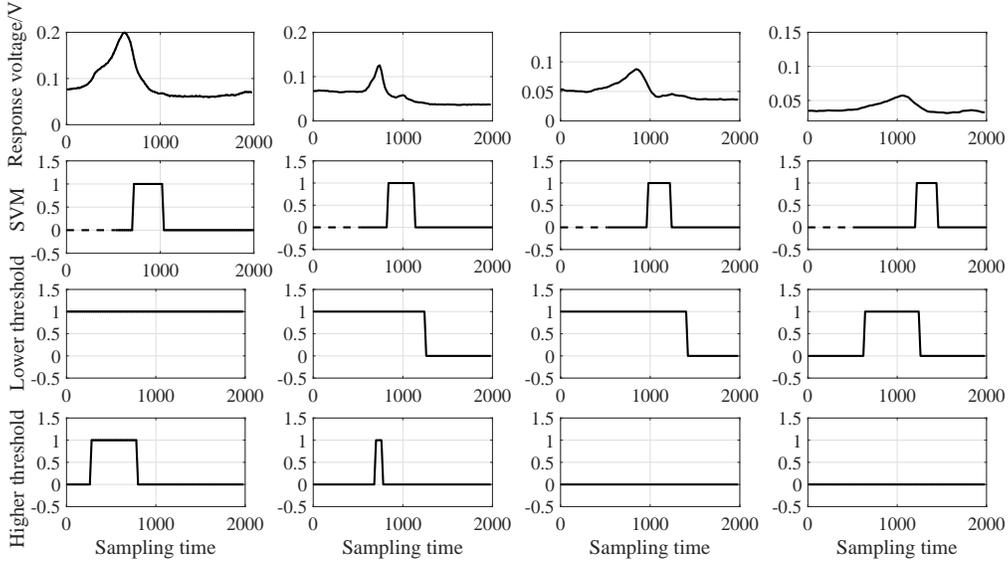}}
	\caption{Test results under different ambient light intensities
		(The ambient light intensity used in the pictures is reduced from left to right.)}
\end{figure*}
%
\subsection{SVM}\label{sec4.2}
SVM is a two-class classification model.
Our input signal $\mathbf{x}_{t}$ is a vector consisting of consecutive sampling points, output signal $Y_{t}$ is whether the corresponding window contains a car passing through the street light,
\begin{equation}\label{split}
Y_{t}=
\begin{cases}
0,\text{no car passes during $T$ sampling periods} \\
1,\text{a car passes during $T$ sampling periods}
\end{cases}.
\end{equation}

We use a training set $D=[\mathbf{x}_t,Y_{t}]$, $t=1,2,3,...,k$ to find a function
\begin{equation}\label{split}
f(\mathbf{x})=sign \{ {\langle\bf w},{\bf x}\rangle+b \},
\end{equation}
to fit the relationship between $\mathbf{x}_t$ and $Y_{t}$, where $\langle \cdot \rangle$ denotes the inner product of two vectors,  ${\bf w}\in \mathbb{R}^{T}$  is the vector of weight parameters and $b$ is a bias factor.

We transform the learning problem into the regularized regression problem by introducing the regularization term of ${\bf w}$,

\begin{equation}\label{split}
\mathbf{w}^{*}={arg min_{{\mathbf w}^{*}}}[\Sigma_{t}\mathcal{L}(f(\mathbf{x}_t),Y_{t})+\alpha\lVert \mathbf{w} \rVert_{2}^{2}],
\end{equation}
where $\mathcal{L}(\cdot)$ is a loss function and $\alpha\geq 0$ is a trade-off parameter that controls the weight of the loss function and the weight of the regularization penalty $ \lVert {\bf w} \rVert_{2}^{2}$.
We use a loss function called Hinge-loss function. Hinge-loss function is used for the ``maximum interval'' classification. The general function expression is,

\begin{equation}\label{split}
\mathcal{L}(f({\mathbf x}_{t}),Y_{t})=max(0,1-Y_{t}f({\mathbf x}_{t})),
\end{equation}
it means that if it is correctly classified, the loss is $0$, otherwise the loss is $1-Y_{t}f({\mathbf x}_{t})$.

As we know, SVM is a linear classifier that can only describe the linear relationship between $\mathbf{x}_t$ and $Y_{t}$. However, our system may be nonlinear. We introduce a nonlinear function $\varphi(\bf x)$ that maps $\mathbf{x}$ to a Reproducing Kernel Hilbert Space. By using the kernel function $K(\mathbf{x}_{i}; \mathbf{x}_{j})=\langle\varphi(\mathbf{x}_{i}); \varphi(\mathbf{x}_{j})\rangle$, we can obtain the optimal solution of (4),
\begin{equation}\label{split}
f^{*}(\mathbf{x})=\Sigma_{t}u_{t}K(\mathbf{x}_{t}; \bf x),
\end{equation}
where $u_{t}$ is obtained by solving the dual problem of (4) and we choose $K(\mathbf{x}_{i}; \mathbf{x}_{j})$ as sigmoid kernel function here.

\section{System Evaluation}\label{sec5}

By using the scaled-down experimental platform shown in Fig. 1 and the algorithm described in Sec. IV, we analyzed and verified the accuracy of the system's detection of traffic flow through experiments.

We adjusted the intensity of light sources to simulate changes in outdoor ambient light intensity. As shown in Fig. 5, the algorithm has strong robustness, and can sense traffic flow information under different ambient light intensities. We adopt threshold detection as a comparison. The threshold is too low to be applied to the case where the light intensity is low, but the false alarm probability is too high. If the threshold is too high, the probability of miss detection is high. The threshold needs to be adjusted in real time according to the ambient light intensity. However, the proposed algorithm has strong adaptability to the environment, as observed from the second row of the figure for varying ambient light intensities. This is because the SVM algorithm has the ability to distinguish peaks and flat lines. The algorithm will generate peaks as long as the car passes through, regardless of the ambient light intensity. The result indicates that the proposed detection system is potentially applicable in a practical outdoor street lighting and sensing environment.


\section{Conclusions}\label{sec6}
In this paper, we developed a novel approach of using LED street lights to sense traffic flow. We built an indoor experimental platform and applied the SVM algorithm to verify the accuracy of this system. In the future work, we hope to apply the system to the outdoor environment, and generalize to other intelligent systems based on traffic flow sensing.



\end{document}